\documentclass[reprint, superscriptaddress, secnumarabic, amssymb, nobibnotes, aps, prl]{revtex4-1}

\usepackage{graphicx}
\usepackage{epstopdf}
\usepackage[T1]{fontenc}
\usepackage{amsbsy}
\usepackage{gensymb}
\usepackage[T1]{fontenc}
\usepackage{amsmath}
\usepackage{amssymb}
\usepackage{bbm}
\usepackage{braket}
\usepackage{xcolor}
\allowdisplaybreaks
\usepackage{graphicx}
\usepackage{natbib}
\usepackage[colorlinks=true]{hyperref}  
\hypersetup{
    bookmarks=true,      
    unicode=false,     
    pdftoolbar=true,        
    pdfmenubar=true,        
    pdffitwindow=false,     
    pdfstartview={FitH},    
    pdftitle={Superconductivity in LixTaS2},    
    pdfauthor={Tarushi Agarwal},      
    pdfsubject={},   
    pdfcreator={},   
    pdfproducer={}, 
    pdfkeywords={} {} {}, 
    pdfnewwindow=true,   
    breaklinks=true,
    colorlinks=true,       
    linkcolor=blue,          
    citecolor=blue,        
    filecolor=magenta,      
    urlcolor=blue        
} 

\usepackage[normalem]{ulem}

\newcommand{\equref}[1]{Eq.~(\ref{#1})}

\newcommand{\figref}[1]{Fig.~\ref{#1}}

\newcommand{\tableref}[1]{Table~\ref{#1}}

\begin{document}

\title{\textrm{Anomalous Magneto-transport and Anisotropic Multigap Superconductivity in Architecturally Misfit Layered System (PbS)$_{1.13}$TaS$_2$}}
\author{Tarushi Agarwal}
\affiliation{Department of Physics, Indian Institute of Science Education and Research Bhopal, Bhopal, 462066, India}
\author{Chandan Patra}
\affiliation{Department of Physics, Indian Institute of Science Education and Research Bhopal, Bhopal, 462066, India}
\author{Poulami Manna}
\affiliation{Department of Physics, Indian Institute of Science Education and Research Bhopal, Bhopal, 462066, India}
\author{Shashank Srivastava}
\affiliation{Department of Physics, Indian Institute of Science Education and Research Bhopal, Bhopal, 462066, India}
\author{Priya Mishra}
\affiliation{Department of Physics, Indian Institute of Science Education and Research Bhopal, Bhopal, 462066, India}
\author{Suhani Sharma}
\affiliation{Department of Physics, Indian Institute of Science Education and Research Bhopal, Bhopal, 462066, India}
\author{Ravi Prakash Singh}
\email[]{rpsingh@iiserb.ac.in}
\affiliation{Department of Physics, Indian Institute of Science Education and Research Bhopal, Bhopal, 462066, India}

\begin{abstract}
Misfit-layered compounds, naturally occurring bulk heterostructures, present a compelling alternative to artificially engineered ones, offering a unique platform for exploring correlated phases and quantum phenomena. This study investigates the magnetotransport and superconducting properties of the misfit compound (PbS)$_{1.13}$TaS$_2$, comprising alternating PbS and 1$H$-TaS$_2$ layers. It exhibits distinctive transport properties, including a prominent planar Hall effect and a four-fold oscillatory Butterfly-shaped anisotropic magnetoresistance (AMR). Moreover, it shows multigap two-dimensional superconductivity with an exceptionally high in-plane upper critical field, exceeding the Pauli limit. The coexistence of unconventional superconductivity and anomalous transport - two distinct quantum phenomena, within the same material, suggests that misfit compounds provide an ideal platform for realizing quantum effects in the two-dimensional limit of bulk crystals. This opens the door to the development of simpler and more efficient quantum devices.
\end{abstract}


\maketitle
\section{INTRODUCTION}
Recent advances in van der Waals (vdW) heterostructures have significantly broadened the exploration of correlated quantum states, providing a versatile platform for the emergence of unconventional superconducting and topological states \cite{vdW_H1,vdW_H2,vdW_H3}. Misfit layer compounds (MLCs), naturally occurring vdW heterostructures \cite{misfit1,misfit2,misfit3} are of particular interest. These materials consist of alternating stacks of monochalcogenides (MX) and transition metal dichalcogenides (TMDs) TX$_2$ layers, typically represented as (MX)$_{1+\delta}$(TX$_2$)$_n$. The MX layer acts as a spacer between two TX$_2$ layers, imparting strong two-dimensionality effects and resulting in highly anisotropic electronic and magnetic behavior. This arrangement provides a unique opportunity to investigate the exotic quantum phenomena of the natural heterostructure of MX/TX$_2$ or monolayer TX$_2$ within a protected bulk environment \cite{SC_misfit1,SC_misfit2}. In monolayer TX$_2$, broken inversion symmetry and strong spin-orbit coupling (SOC) result in pronounced spin splitting and drastically alter the band structure compared to their bulk counterparts \cite{SOC_TMD1, SOC_TMD2}. Notably, similar SOC effects can be achieved in bulk misfit compounds due to decoupled TX$_2$ layers. Combination of reduced dimensionality and SOC in MLCs can lead to a range of physical phenomena, including the realization of Ising/unconventional superconductivity in bulk superlattices, diverse magnetic behaviors, and topologically driven anomalous transport properties \cite{ising_misfit1,ising_misfit2,SOC_misfit1, SOC_misfit2, SOC_misfit3}. 

Despite the promising potential of MLCs to exhibit anomalous transport phenomena arising from their inherent heterostructure/monolayer character and strong SOC, their magnetotransport properties remain largely unexplored \cite{LaNSe3,SnNbS3}. The reduced symmetry and modified band structure of these bulk superlattices can give rise to distinct electronic phenomena, such as anisotropic magnetoresistance and nonlinear Hall effects, which are sensitive to interlayer coupling and interfacial effects \cite{SnNbS3, PbTaSe2}. The presence of SOC can further induce spin-dependent transport phenomena, which is recognized as an experimental probe for revealing exotic quantum states \cite{SOC_TM1, SOC_TM2, SOC_TM3, transport_TM1, transport_TM2}. Exploring these transport properties offers new insights into correlated electronic states in MLCs and could facilitate the development of advanced quantum devices with natural heterostructures/ encapsulated TMD layers.   

This work presents investigation of the transport and superconducting properties of (PbS)$_{1.13}$TaS$_2$, a misfit compound composed of the alternating stacking of PbS and 1$H$-TaS$_2$ layers. The PbS layer significantly decouples the TaS$_2$ layers, enhancing their two-dimensionality. This enhanced two-dimensionality is evident in the elevated superconducting transition temperature ($T_c$) 3.05(5) K, which is comparable to the few-layer TaS$_2$ \cite{TaS2thin_SC, TaS2_ising}. We have observed multigap 2D superconductivity in (PbS)$_{1.13}$TaS$_2$ with a high in-plane upper critical field value, violating the Pauli limit, suggesting the possible existence of Ising-like superconductivity. Furthermore, we present the first observation of anomalous magnetotransport in this misfit compound. The normal-state magnetotransport measurements reveal a butterfly-shaped four-fold anisotropic magnetoresistance (AMR), planar Hall effect (PHE), and non-saturating linear MR in (PbS)$_{1.13}$TaS$_2$ single crystal, mainly originating from strong SOC effects. These transport properties indicate the non-trivial band topology of (PbS)$_{1.13}$TaS$_2$ misfit compound. These novel findings, encompassing both superconducting and anomalous magnetotransport properties, offer new avenues for probing 2D phenomena within a bulk vdW superlattice, overcoming the limitations of conventional heterostructures.

\section{RESULTS AND DISCUSSION}
\begin{figure*}[tp]
\centering
\includegraphics[width=0.99\textwidth]{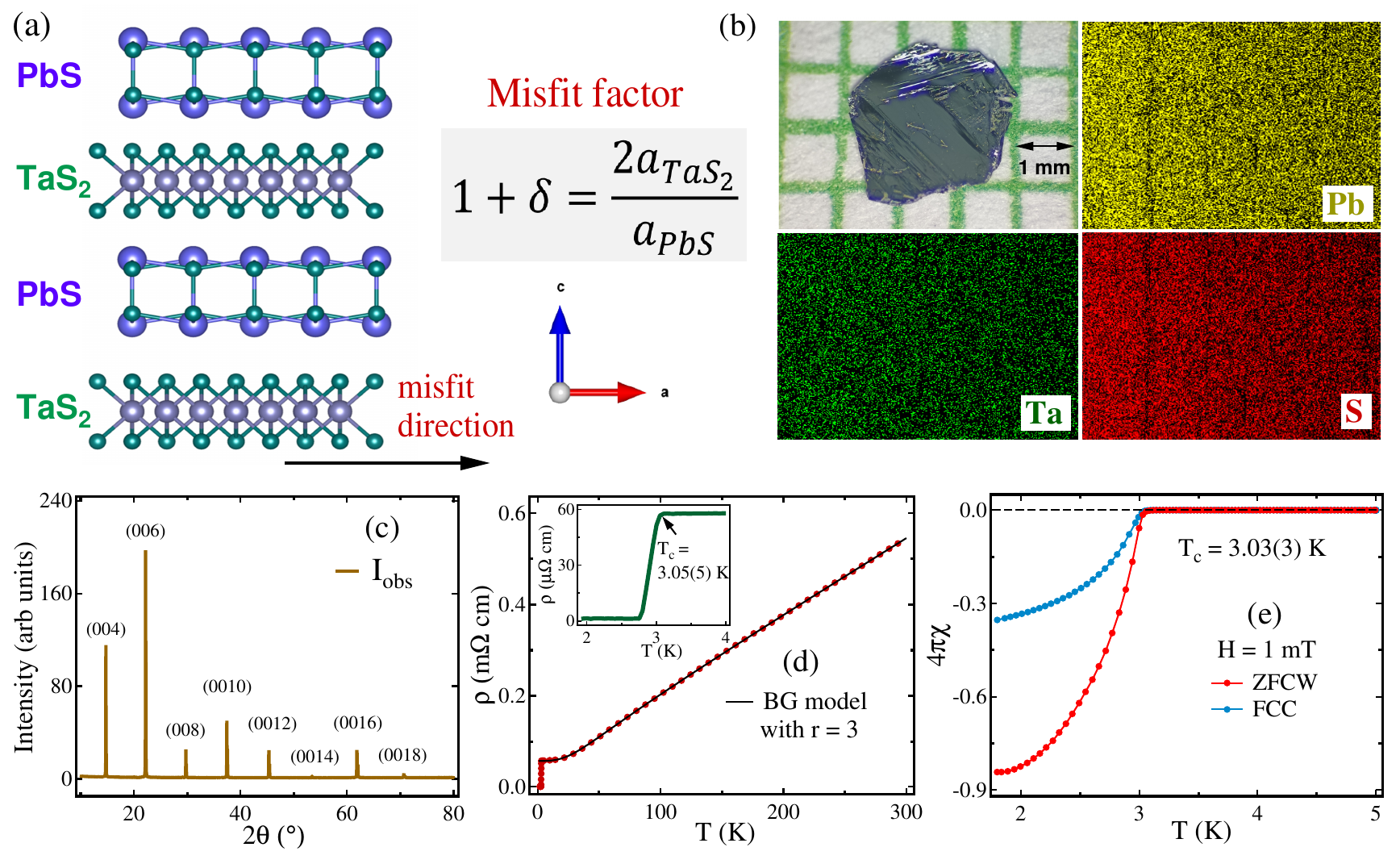}
\caption{\label{Fig1}(a) The crystal structure of misfit compound (PbS)$_{1.13}$TaS$_2$ formed by alternate stacking of PbS and TaS$_2$ layers. (b) The microscopic image of a grown crystal and EDS elemental mapping, indicating the uniform distribution of elements over the surface of (PbS)$_{1.13}$TaS$_2$ crystal. (c) Single-crystal XRD pattern of (PbS)$_{1.13}$TaS$_2$. (d) Zero field temperature dependence of resistivity. The black line indicates the fit using the BG model. The inset demonstrates the transition into the superconducting state at 3.05(5) K. (e) Superconducting diamagnetic behavior in (PbS)$_{1.13}$TaS$_2$ by low-temperature magnetization measurement.}    
\end{figure*}

The crystal structure of (PbS)$_{1.13}$TaS$_2$ is presented in \figref{Fig1}(a) with the alternate stacking of the layers of PbS and 1$H$-TaS$_2$. \figref{Fig1}(b) shows the microscopic image of the grown crystal of (PbS)$_{1.13}$TaS$_2$. The elemental mapping in the single crystal was performed using energy-dispersive X-ray spectroscopy (EDS), confirming the uniform distribution of Pb, Ta, and S elements in (PbS)$_{1.13}$TaS$_2$.
\begin{table}[b]
\centering
\caption{The lattice parameters of (PbS)$_{1.13}$TaS$_2$}
\label{table1}
\setlength{\tabcolsep}{11.5pt}
\begin{tabular}{p{1.3cm}p{0.87cm}p{0.7cm}p{1.0cm}p{0.7cm}}\hline
System & $a~(\text{\AA})$ & $b~(\text{\AA})$ & $c~(\text{\AA})$ & Space group  \\
\hline
\hline
PbS & 5.825 & 5.779 & 23.96 & \textit{Fm2m}\\
TaS$_2$ & 3.304 & 5.779 & 23.96 & \textit{Fm2m}\\
(PbS)$_{1.13}$TaS$_2$ & 23.297(5) & 5.743(7) & 23.994(2) & \textit{Pmmm}\\
\hline
\end{tabular}
\end{table}
The single crystal XRD patterns of (PbS)$_{1.13}$TaS$_2$ are shown in \figref{Fig1}(c). The sharp diffraction peaks with only [00$l$] reflections in single crystal XRD indicate $c$-axis oriented crystals with good quality. The crystal structure of (PbS)$_{1.13}$TaS$_2$ supercell was determined by profile fitting of the powder XRD pattern, obtaining an orthorhombic structure with space group \textit{Pmmm} (see supplementary Fig.S1). The obtained lattice parameters are listed in \tableref{table1}, which is consistent with previously reported values \cite{DFT_PbTS3}. 

\section{Superconducting properties of $\text{(PbS)$_{1.13}$TaS$_2$}$}
Low-temperature electrical resistivity measurements evidenced superconductivity in bulk (PbS)$_{1.13}$TaS$_2$ single crystal with the zero drop in resistivity at the onset temperature $T_{c}^{onset}$ = 3.05(5) K, as shown in the inset of \figref{Fig1}(d). The observed $T_c$ is higher compared to bulk 2$H$-TaS$_2$ ($\sim$ 0.8 K) but consistent with the value of monolayer TaS$_2$ \cite{TaS2_ising,TaS2bulk_SC,monolayer_tas2}. Above the transition temperature, the linear response of resistivity (dR/dT > 0) indicates a metallic normal state with a residual resistivity ratio (RRR) of 9.3 which is explained by the Bloch-Grüneisen (BG) model (see details in the Supplement).
\begin{figure*}[th!]
\centering
\includegraphics[width=0.7\textwidth]{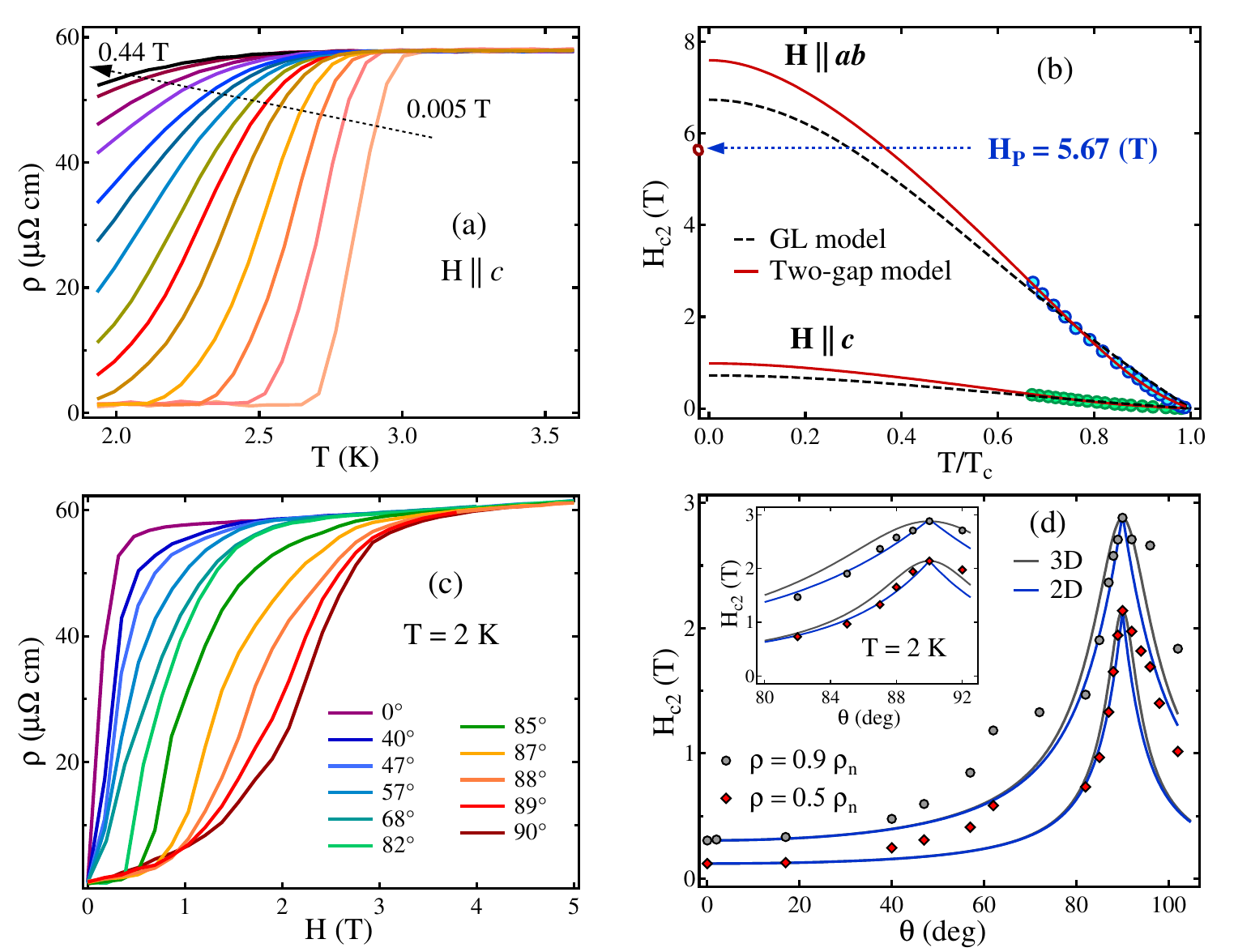}
\caption{\label{Fig2}(a) Temperature dependence of resistivity in the superconducting state under different fields, applied along $H || c$ directions. (b) The deduced $H_{c2}$ values as a function of $T$ for the bulk (PbS)$_{1.13}$TaS$_2$ compound. The dashed black and solid red lines indicate the fit by GL and two-gap model, respectively. (c) Resistivity as a function of $H$, measured at different angles at $T$ = 2 K. (d) The extracted anisotropic variation of $H_{c2}$ as a function of $\theta$. The solid lines show the fits by 3D AGL and 2D Tinkham models. A zoomed view of the data is shown in the inset across $90^\circ$.}  
\end{figure*}

Superconductivity in (PbS)$_{1.13}$TaS$_2$ is also confirmed by the temperature dependence of magnetization and zero-field specific heat measurement (see supplementary Fig.S4). Strong diamagnetic behavior of magnetic susceptibility indicates the superconducting transition with the onset temperature $T_c$ = 3.03(2) K, similar to the value obtained by resistivity. Magnetization measurement was performed in the zero-field cooled warming (ZFCW) and field cooled cooling (FCC) modes at $H$ = 1 mT, applied along the $c$-axis, as shown in \figref{Fig1}(e). In addition, the superconducting gap parameters are evaluated by fitting the electronic specific heat data below $T_c$, suggesting the two-gap superconductivity in (PbS)$_{1.13}$TaS$_2$ (see supplementary Fig.S4). The obtained gap values $\Delta_0$ are 0.43(4) meV and 0.20(8) meV, where the dominating larger gap value closely aligns that of monolayer 1$H$-TaS$_2$ \cite{gap_TaS2_1,gap_TaS2_2}, manifesting the monolayer effect of TaS$_2$ layer within the bulk (PbS)$_{1.13}$TaS$_2$ superlattice.

The upper critical field values ($H_{c2}$) are extracted from the temperature-dependent resistivity measurement at different fields, parallel ($H || ab$) and perpendicular ($H || c$) to the crystal plane, shown in \figref{Fig2}(a). The $T_c$ decreases with the increase in magnetic field, and accordingly, we plotted the temperature dependence of $H_{c2}$ along both directions. We observed that superconductivity is more susceptible when the field is applied perpendicular to the plane of the layers. The $H_{c2}(T)$ values were determined by considering $\rho$ = 0.9$\rho_n$, where $\rho_n$ is the normal state resistivity. The $T$ dependence of $H_{c2}$ is presented in \figref{Fig2}(b) with a positive curvature near $T_c$ for both $H || ab$ and $H || c$ directions. This kind of behavior has been observed in MgB$_2$, some iron-based superconductors, and misfit compounds \cite{2gap1,2gap2,2gap3}, indicative of multigap superconductivity, and cannot be described by the Ginzburg-Landau (GL) model. A two-band model explains the dependence $T$ on $H_{c2}$ (see details in the supplementary), indicated by the red solid line in \figref{Fig2}(b)  and the determined values are: $H_{c2}^{|| ab}$(0) = 7.37(8) T, and $H_{c2}^{||c}$(0) = 0.98(3) T. The different values of the critical fields for different orientations signify the anisotropic nature of the superconductivity, and the estimated anisotropy parameter is $\Gamma$ = ($H_{c2}^{|| ab}$(0)/ $H_{c2}^{||c}$(0)) = 7.5, coordinating with values observed in other misfit compounds \cite{PbNbSe3}. Meanwhile, the estimated $H_{c2}^{|| ab}$(0) value for bulk (PbS)$_{1.13}$TaS$_2$ crystal exceeds the Pauli limiting field, calculated as $H_P=1.86$(T K$^{-1}$) $T_{c}$, yielding 5.67 T. This violation of the Pauli limit suggests the dominance of strong spin-orbit coupling, arising from existing significant decoupled 1$H$-TaS$_2$ layers \cite{TaS2_ising}. It highlights that the misfit compound (PbS)$_{1.13}$TaS$_2$ can be a promising candidate to investigate Ising-like superconductivity in bulk systems.
\begin{figure*}[t!]
    \centering
    \includegraphics[width=\textwidth]{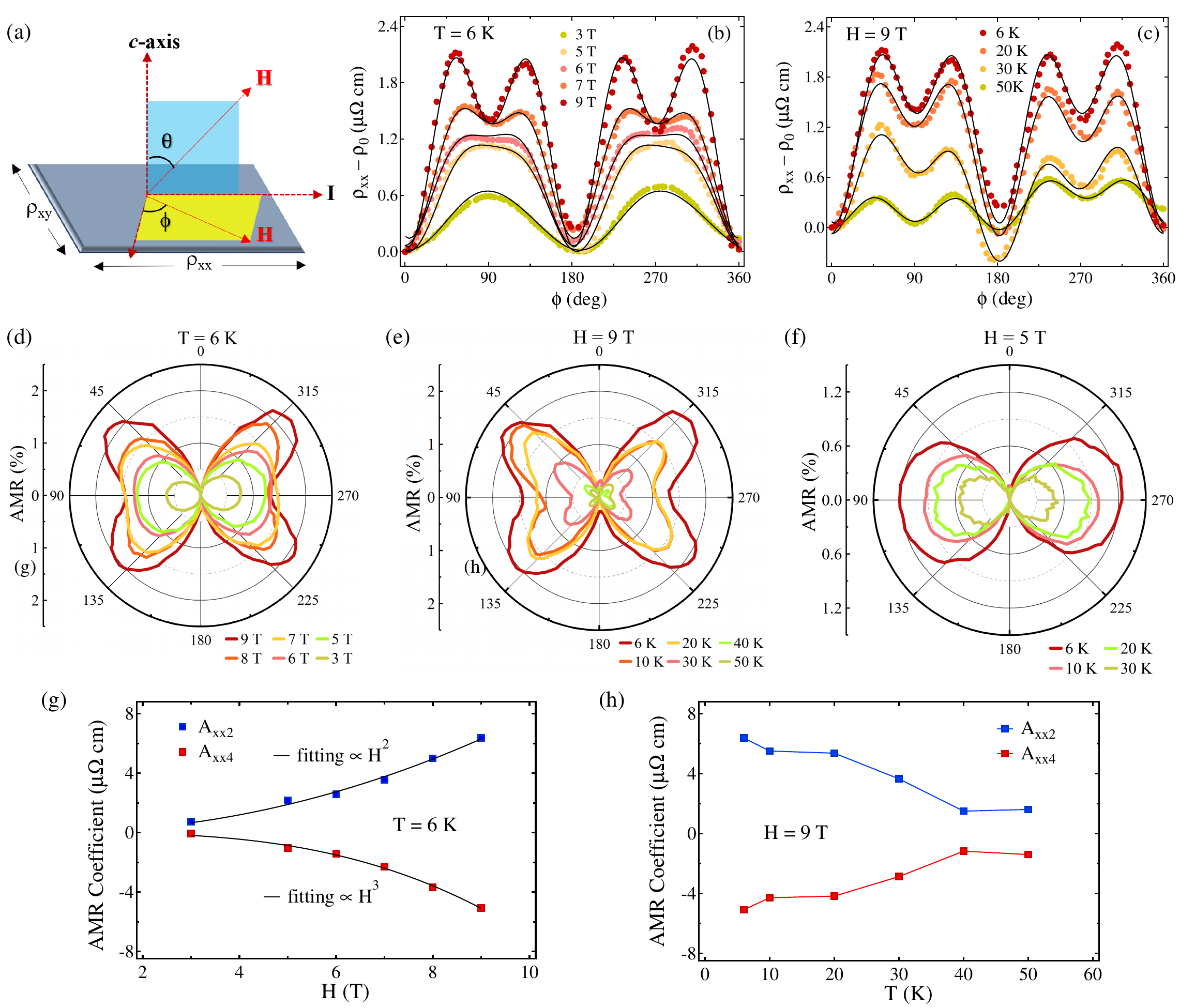}
    \caption{\label{Fig3}(a) Schematic of $\phi$, and $\theta$ variation of applied magnetic field. Four-fold symmetric AMR and the corresponding fitting curves (indicated by blacklines) under (b) different magnetic fields at $T$ = 6 K, and (c) different temperatures at $H$ = 9 T. (d) and (e) show the Polar plot of AMR ratio demonstrating the butterfly AMR effect. (f) AMR polar plot with only two-fold symmetry dependencies, measured at $H$ = 5 T. The field and temperature dependence of AMR two-fold and four-fold coefficients are shown in (g) and (h), respectively. The solid black line indicates the fitting.}    
\end{figure*}

Further, to determine the nature of superconductivity in (PbS)$_{1.13}$TaS$_2$, we measured the out-of-plane angular variation of H$_{c2}$. \figref{Fig2}(c) represents the field dependence of resistivity at different angles $\theta$ at a fixed temperature $T$ = 2 K, where $\theta$ is the angle between the direction of the applied field and the normal of the crystal plane. The H$_{c2}$ values are determined as the magnetic field at which the resistivity value drops to 90$\%$ and 50$\%$ of the normal state resistivity. The obtained angular variation of $H_{c2}$ is summarized in \figref{Fig2}(d), explained by using 3D anisotropic GL and 2D Tinkham models \cite{3dmodel,2dmodel}. 
Using these two models, we found that the behavior of $H_{c2}(\theta)$ aligns with the 2D Tinkham model, similarly observed for other misfit layered superconductors \cite{2d_misfit1,2d_misfit2}. This indicates significant decoupling of two adjacent TaS$_2$ layers due to the existing non-superconducting spacer PbS layer, effectively inducing 2D superconductivity in bulk (PbS)$_{1.13}$TaS$_2$. Thus, MLCs serve as robust platforms for probing a range of 2D properties within bulk vdW superlattices.

\subsection{Magneto-transport properties}
\begin{figure*}[t!]
    \centering
    \includegraphics[width=\textwidth]{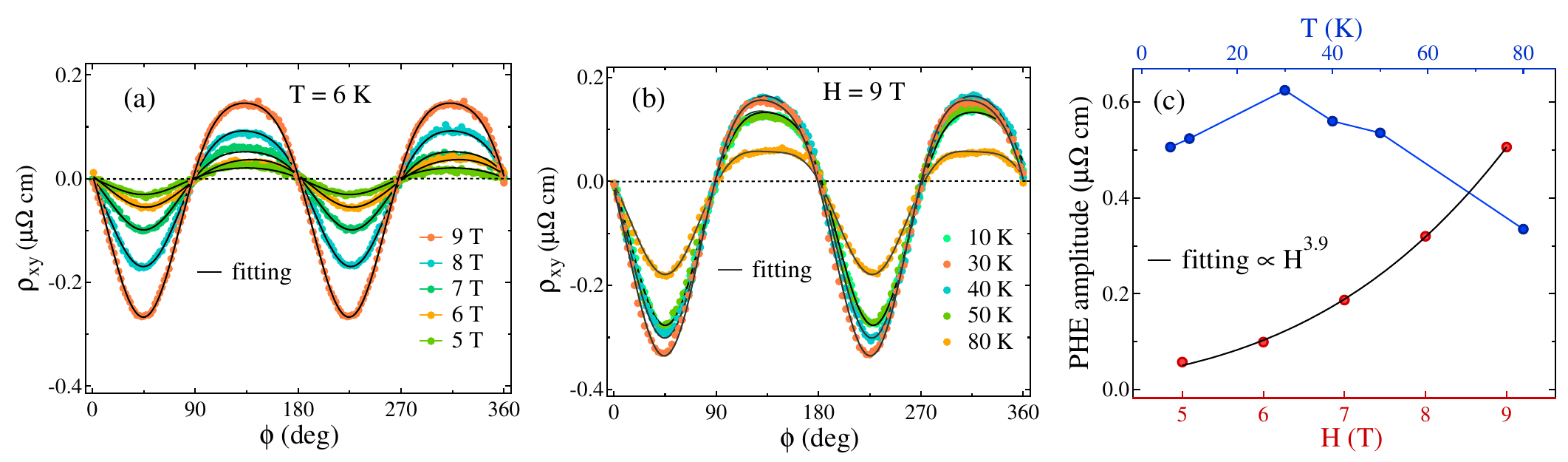}
    \caption{\label{Fig4}Angular variation of $\rho_{xy}$, measured under (a) different magnetic fields and (b) temperatures. The black line indicates the fitted curve using \equref{eq4}. (c) The field and temperature dependence of the PHE amplitude are represented with red and blue markers, respectively.} 
\end{figure*}
The magneto-transport properties of (PbS)$_{1.13}$TaS$_2$ are studied by measuring the in-plane angle-dependence of longitudinal ($\rho_{xx}$) and transverse resistivity ($\rho_{xy}$). The PHE and AMR are key transport signatures that have been extensively investigated in several quantum materials and are frequently associated with chiral anomaly and the presence of topological phases in non-magnetic systems \cite{WSM_PHE, WSM1,WSM2}. These properties strongly depend on the position of Dirac or Weyl points related to the Fermi surface. According to conventional PHE and AMR behavior, two-fold oscillations have been observed in several topological materials, in which the resulting transverse and longitudinal resistivity varies as \cite{PHE_formula}: 
\begin{gather}
\rho_{xy}^{PHE} = -\Delta \rho_{xy}^{chiral}\sin{\phi}\cos{\phi}\\ \rho_{xx}^{AMR} = \rho_{0}-\Delta\rho_{xx}^{chiral}\cos^2{\phi}. 
\label{eq2}
\end{gather}
Here, $\Delta \rho_{xy}^{chiral}$ and $\Delta \rho_{xx}^{chiral}$ represent chiral anomaly-induced oscillatory transverse and longitudinal resistivity, respectively. In contrast to this feature, four-fold oscillatory AMR is observed in (PbS)$_{1.13}$TaS$_2$ under various applied magnetic fields and temperatures. \figref{Fig3}(a) shows a schematic diagram of the measurement configuration of AMR and PHE at the angle position $\phi$, which is the angle between the current direction and the in-plane applied field. The angular variation of $(\rho_{xx}-\rho_0)$ for different fields and temperatures is presented in \figref{Fig3}(b) and \figref{Fig3}(c), where $\rho_0$ is the resistivity when the magnetic field is perpendicular to the current flow. To define these oscillations, $\rho_{xx}$ can be expressed as \cite{fit_eq}:
\begin{multline}
   \rho_{xx}^{AMR} = A_{xx0}+A_{xx1}\cos{(\phi-\phi_{U})}+A_{xx2}\cos^2{(\phi-\phi^{'})}\\+A_{xx4}\cos^4{(\phi-\phi^{'}).}
     \label{eq3}     
\end{multline}
Here, $A_{xx2}$ and $A_{xx4}$ give the coefficients for two-fold and four-fold oscillations of AMR. Moreover, to better fit the AMR, an additional second term was added in \equref{eq3}, indicating a unidirectional component, that is, $\rho(0^{\circ})\neq \rho(180^{\circ})$ and $A_{xx1}$ is the coefficient of the unidirectional term. The different AMR curves were very well fitted with \equref{eq3}, as illustrated in \figref{Fig3}(b) and \figref{Fig3}(c) by black solid lines. The angular dependence of the AMR ratio is also presented on a polar plot, where the AMR ratio is defined as AMR ($\%$) = $(\rho_{xx}-\rho_0)/\rho_0)$. \figref{Fig3}(d) and \figref{Fig3}(e) show the butterfly AMR, measured at $T$ = 6 K for various magnetic field strengths and at $H$ = 9 T at different temperatures. 

The butterfly pattern in AMR has been observed across various magnetic systems, mainly attributed to anisotropic magnetic scattering, relaxation time anisotropy and higher-order perturbation of SOC \cite{butterfly_amr1,butterfly_amr2,butterfly_amr3,butterfly_amr4,butterfly_amr5}. In contrast, this feature is rarely seen in non-magnetic systems, where it is predominantly linked to signatures of topologically protected states \cite{ZrSiS, ZrTe5,MoAs2}. Furthermore, we found that the four-fold symmetry of AMR merged into two-fold oscillations with decreasing field as shown in \figref{Fig3}(b), observed for (PbS)$_{1.13}$TaS$_2$. \figref{Fig3}(f) illustrates the polar plot of the AMR ratio at $H$ = 5 T, where two-fold oscillations were detected with decreasing amplitudes as we increased the temperature. The fourfold symmetric AMR and the evolution in AMR symmetry with a magnetic field strongly signal the dominant effect of spin-orbit coupling on electron scattering dynamics in (PbS)$_{1.13}$TaS$_2$. In-plane lattice anisotropy can also play a role in hosting the anisotropic carrier scattering. However, further information regarding the Fermi surface topology is required to ascertain any potential non-trivial origins.  

By fitting the different AMR curves with \equref{eq3}, the field and temperature dependence of the two- and four-fold coefficients were extracted. At $T$ = 6 K, the two-fold coefficient exhibits a quadratic ($A_{xx2} \propto H^{2}$) field dependence, while the four-fold coefficient follows a nearly cubic ($A_{xx4} \propto H^{3}$) relation (\figref{Fig3}(g)), suggesting that the four-fold components diminish more rapidly as the magnetic field decreases. Moreover, both coefficients exhibit an almost similar temperature dependence, represented by \figref{Fig2}(h). 
\begin{figure*}[t!]
\centering
\includegraphics[width=\textwidth]{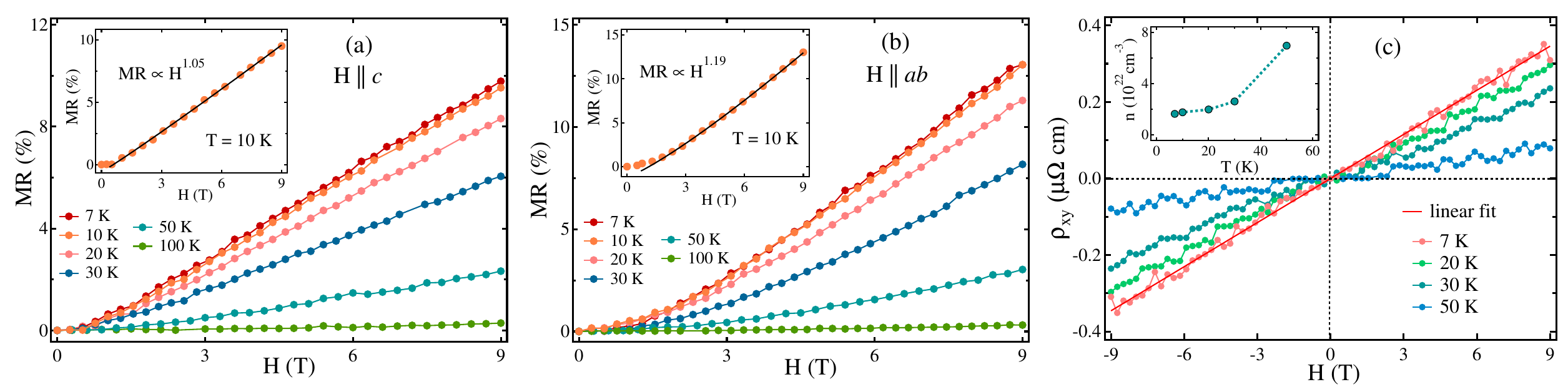}
\caption{\label{Fig5}The observed positive non-saturating linear MR, particularly over the high field range. (a), and (b) show the MR behavior for $H||c$, and $H||ab$ field directions, respectively. The insets present the power law fitting at $T$ = 10 K. (c) Linear variation of Hall resistivity of (PbS)$_{1.13}$TaS$_2$ crystal with positive slope, measured at different temperatures. The inset shows the hole concentrations at different temperatures.}
\end{figure*}

In further exploration of transverse resistivity, we observed PHE in (PbS)$_{1.13}$TaS$_2$ single crystal, measured under various fields and temperatures with a rotating magnetic field in the $ab$-plane. PHE shows two-fold oscillations with a finite value for all field directions except the parallel and perpendicular field directions. The observed PHE behavior is defined by the equation as follows \cite{fit_eq}: 
\begin{multline}
\rho_{xy}^{PHE} = C_{xy0}+C_{xy1}\cos^4{(\phi)}+C_{xy2}\sin{(\phi-\phi^{'})}\\\cos{(\phi-\phi^{'}).}
\label{eq4}
\end{multline}
We have added an extra term in \equref{eq4}, indicating the additional AMR contribution. Our data is well defined by \equref{eq4}, as shown in \figref{Fig4}(c), and \figref{Fig4}(d). The magnetic field and temperature dependence of the PHE amplitude $C_{xy2}$, deduced from the fitting, is shown in \figref{Fig4}(e). The value of $C_{xy2}$ increases as $H$ increases following the $H^4$ relation, different from the quadratic behavior expected for chiral anomaly-driven PHE. The values of $C_{xy2}$ as a function of $T$ reveal that up to 80 K there is no significant suppression in the value of $C_{xy2}$ with increasing temperature.  Furthermore, the amplitude $C_{xy2}$ initially increases to $T$ = 30 K before decreasing with a further rise in temperature, showing unusual behavior compared to other systems reported in\cite{WSM2,PHE1,PHE2}.   

Transport properties of (PbS)$_{1.13}$TaS$_2$ crystal are continued by measuring linear MR behavior, illustrated in \figref{Fig5}(a) and \figref{Fig5}(b). MR was measured in two different orientations, particularly when the magnetic field is applied in the in-plane ($H || ab$) and out-of-plane ($H ||c$) directions. MR can be defined as MR = [$\rho_{xx}(B)-\rho_{xx}(0)$]/$\rho_{xx}(0)$, where $\rho_{xx}(B)$ and $\rho_{xx}(0)$ are the longitudinal resistivity measured with and without applied field, respectively. A large, non-saturating positive MR is obtained along both directions of the applied field across all measured temperatures. Specifically, for the $H$||$ab$ orientation, a remarkable MR value of 13$\%$ is observed. For both field directions, the MR exhibits a quadratic field dependence at lower fields, transitioning to a linear relationship at higher fields. The insets of \figref{Fig5} demonstrate the fit of MR data using a power law MR $\propto B^n$ for $T$ = 10 K with $n$ varying between 1.05 and 1.2. Non-saturating linear MR behavior has been observed in various Dirac and Weyl semimetals, and topological insulators, where several mechanisms have been proposed to explain this phenomenon, including linear band dispersion, complex Fermi surfaces, and disorder effect \cite{MR1,MR2,MR3}.

Further, we measured normal Hall resistivity by sweeping the fields up to 9 T to calculate the carrier density of (PbS)$_{1.13}$TaS$_2$. The linear variation of Hall resistivity $\rho_{xy}$ with varying magnetic field, applied along the $c$ direction, is shown in \figref{Fig5}(c). By fitting a simple straight-line equation, slope values are estimated, which give the Hall coefficient ($R_H$) at a particular temperature. The positive values obtained of $R_H$ denote the dominant hole-type charge carriers in bulk (PbS)$_{1.13}$TaS$_2$. The carrier density can be calculated using the formula $R_H = 1/ne$, where $e$ is the charge of an electron. The inset of \figref{Fig5}(c) shows the hole density values at various temperatures, which are on the order of $10^{22}$ holes/cm$^{3}$. These values are higher than the carrier density of bulk 2\textit{H}-TaS$_2$ \cite{PdTaS2} and closely match those of intercalated 2\textit{H}-TaS$_2$, signifying charge transfer from PbS to the TaS$_2$ layer \cite{LiTaS2,CuTaS2,DFT_PbTS3}. 

\section{Discussion}
Our findings underscore the dominant role of decoupled 1$H$-TaS$_2$ layers and their strong spin-orbit coupling (SOC) in governing the electronic properties of the misfit compound (PbS)$_{1.13}$TaS$_2$. The observed superconducting transition at $T_c$ = 3.05 (5) K, closely mirrored that of monolayer 1$H$-TaS$_2$, provides compelling evidence for this decoupling. The high in-plane $H_{c2}$(0), further support the presence of Ising-like superconductivity in our compound, opening avenues for exploring 2D unconventional superconducting phases within the bulk superlattice.  

Further four-fold symmetric AMR and PHE in (PbS)$_{1.13}$TaS$_2$ strongly suggest intriguing underlying electronic properties. The pronounced directional dependence of the resistivity in the four-fold AMR highlights its sensitivity to strong SOC effects. In addition, this behavior could be attributed to the inherent crystalline anisotropy or the non-trivial band characteristics of the system, potentially arising from the interplay between these factors and SOC. 
The two-fold PHE in (PbS)$_{1.13}$TaS$_2$ further emphasizes this complexity. The determined $H^4$ dependence of the PHE amplitude and the observed positive linear MR rule out the chiral anomaly as its origin. In addition, in other mechanisms, the orbital magnetoresistance and spin-scattering effect can be a possible origin of PHE in (PbS)$_{1.13}$TaS$_2$. We believe that the large strength of the Ising SOC \cite{SOC_misfit2} due to the decoupled TaS$_2$ layers strongly supports the dominant orbital contribution in our sample and induces the PHE response \cite{orb_PHE1,orb_PHE2}. Consequently, such MLC materials can host the in-plane components of the band geometric quantities, which is a way to probe topological band structures in such quasi-2D materials. These transport properties position (PbS)$_{1.13}$TaS$_2$ as a possible topological material, warranting further in-depth theoretical investigation. 
Future research directions involve considering (PbS)$_{1.13}$TaS$_2$ as a heterostructure comprising a topological crystalline insulator (TCI) PbS layer and an Ising superconducting 1$H$-TaS$_2$ layer with weak interlayer coupling \cite{PbS,TaS2_ising}. By systematically tuning the interlayer coupling, (PbS)$_{1.13}$TaS$_2$ could emerge as a promising candidate to realize topological superconductivity \cite{TSC1,TSC2,TSC3}. Investigating the layer-dependent properties of (PbS)$_{1.13}$TaS$_2$, which could enhance SOC effects and induce more prominent 2D phenomena, is crucial to further understanding its intriguing properties.

\section{Conclusion}
In summary, we have successfully synthesized the single crystals of the misfit layered compound (PbS)$_{1.13}$TaS$_2$. The (PbS)$_{1.13}$TaS$_2$, formed by the 2D stack of 1$H$-TaS$_2$ layer combined with the intervening PbS layer, offers an ideal route to study the exotic quantum properties of monolayer 1$H$-TaS$_2$ in a protected environment. Low-temperature measurements confirm the multigap superconductivity in (PbS)$_{1.13}$TaS$_2$ below the transition temperature of $T_c$ = 3.05(5) K, aligned with monolayer 1$H$-TaS$_2$. The high in-plane $H_{c2}$ value, exceeding the Pauli limit, indicates the dominance of the ising-SOC effect in bulk (PbS)$_{1.13}$TaS$_2$. In key results of this study, we observed butterfly-shaped AMR and two-fold PHE characteristics of the normal state of (PbS)$_{1.13}$TaS$_2$ crystal. These distinctive magnetotransport features arise from the dominant influence of spin-orbit coupling (SOC) on electron scattering within (PbS)$_{1.13}$TaS$_2$. Furthermore, the combination of decoupled 1$H$-TaS$_2$ layers, strong SOC, and broken inversion symmetry in this system could give rise to non-trivial band geometric quantities in the presence of an in-plane magnetic field, a characteristic often observed in quasi-2D materials. However, a comprehensive understanding of these behaviors necessitates further spectroscopic measurements and detailed theoretical modeling to fully elucidate the Fermi surface topology of (PbS)$_{1.13}$TaS$_2$. Our findings offer a unique opportunity to delve deeper into the microscopic electronic states of misfit layered compounds, enabling the study of monolayer properties within a bulk environment, and paving the way for the development of highly efficient devices based on these self-assembled heterostructures.

\section{Experimental Details}
Single crystals of misfit compound (PbS)$_{1.13}$TaS$_2$ were prepared using the chemical vapour transport (CVT) method using iodine as a transport agent. First, the polycrystal sample was synthesized by mixing the raw elements in 1.13$\colon$1$\colon$3.13 ratio and heated at 1173 K for four days, followed by ice-water quenching. The resulting mixture was sealed in an evacuated quartz tube and placed in a two-zone furnace, where the mixture was heated in a temperature gradient of 1203 K-1123 K for ten days, and large crystals were obtained with dimensions of 4$\times$4$\times$0.5 mm. The powder and single crystal XRD pattern of (PbS)$_{1.13}$TaS$_2$ were recorded on a Panalytical diffractometer equipped with Cu $K_\alpha$ radiation. The Laue diffraction and EDX spectra of a single crystal were recorded using the Photonic–Science Laue camera system and scanning electron microscope (SEM), respectively. Transport and specific heat measurements were performed using the Quantum Design Physical Property Measurement System (PPMS). A standard four-probe method has been used for all transport measurements. Magnetization measurements down to 1.8 K were carried out in a magnetic property measurement system (MPMS).

\section{Acknowledgments}

T. Agarwal acknowledges the Department of Science and Technology (DST), Government of India, for providing the SRF fellowship (Award No. DST/INSPIRE/03/2021/002666). R. P. S. acknowledges the SERB, Government of India, for the Core Research Grant No. CRG/2023/000817. We would like to thank Dr. Ganesh Ji Omar for valuable discussions.

\section{Supplementary information}

\subsection{Sample characterization}
\begin{figure*}
    \centering
    \includegraphics[width=0.8\textwidth]{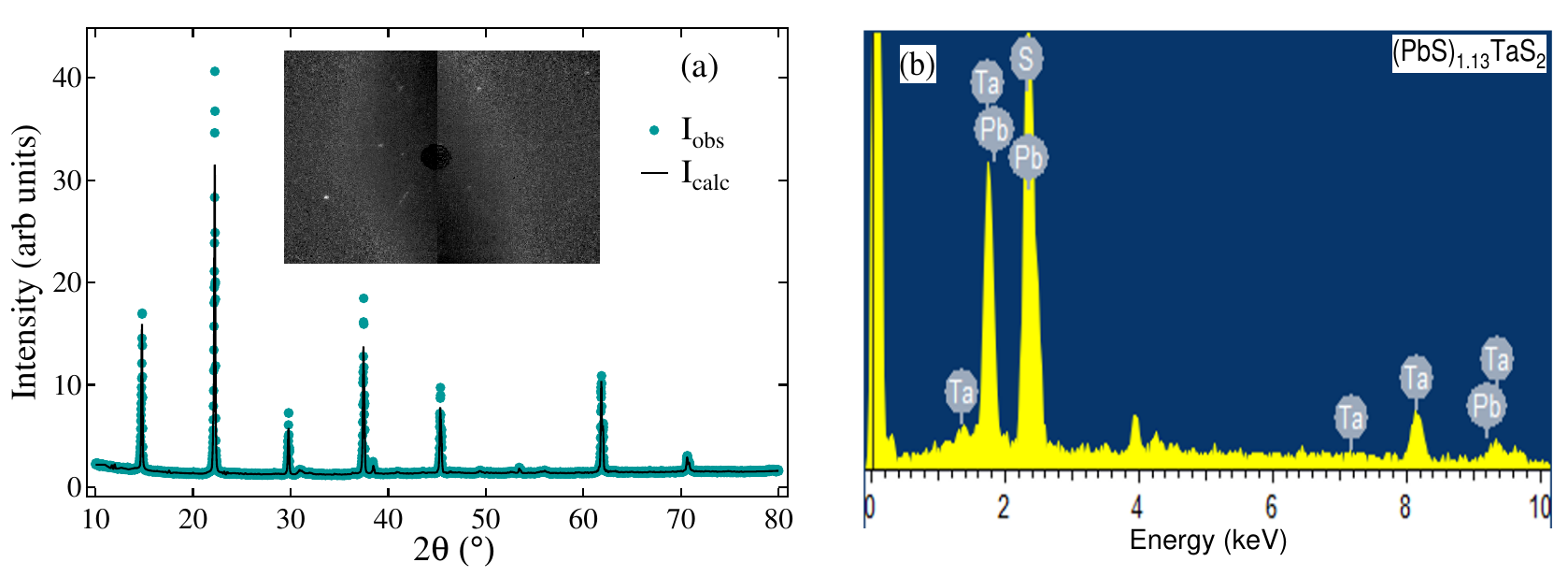}
    \caption{\label{fig1}(a) X-ray diffraction pattern of polycrystal powder of (PbS)$_{1.13}$TaS$_2$. The black line indicates the calculated intensity by profile fitting. The Laue diffraction pattern is shown in the inset. (b) EDS spectra of (PbS)$_{1.13}$TaS$_2$ single crystal.}   
\end{figure*}

The structure of (PbS)$_{1.13}$TaS$_2$ compound is determined by profile fitting of powder XRD pattern, without considering the atom positions, shown in \figref{fig1}(a). The inset shows the Laue diffraction pattern of (PbS)$_{1.13}$TaS$_2$ single crystal, displaying hexagonal symmetry.  The chemical composition of (PbS)$_{1.13}$TaS$_2$ is confirmed by performing energy dispersive x-ray spectroscopy (EDS), yielding Pb$_{0.9}$: Ta$_1$: S$_{3.02}$ which gives approximately PbTaS$_3$ composition and \figref{fig1}(b) presents the corresponding EDS spectra.  

\subsection{Magnetization measurement}
\begin{figure*}
    \centering
    \includegraphics[width=\textwidth]{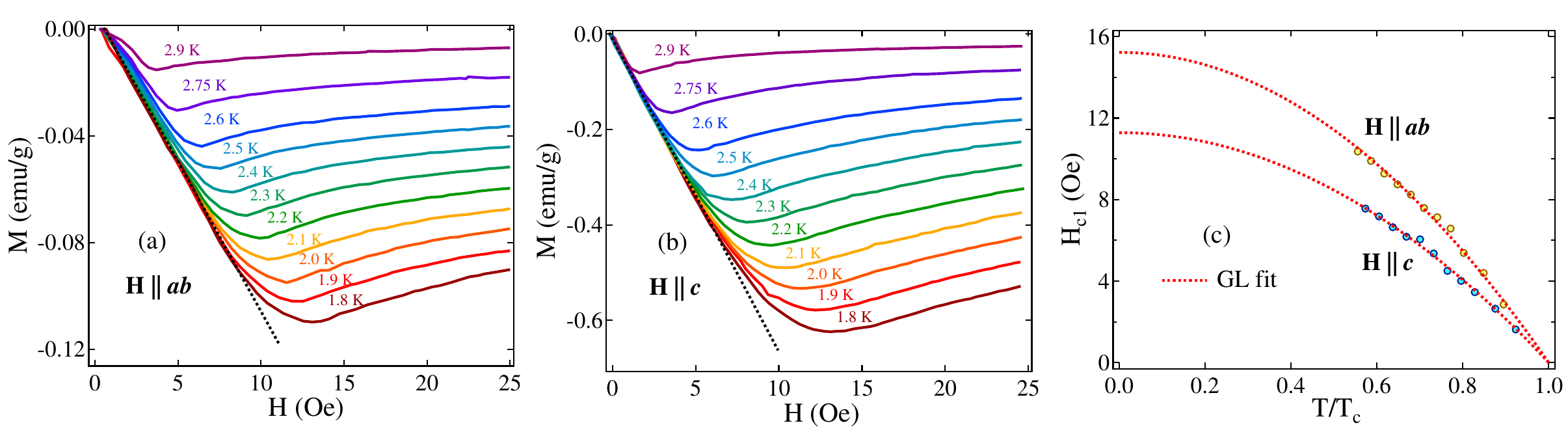}
    \caption{\label{fig2}Low field magnetization curves at different temperatures for (a) $H || ab$, and (b) $H || c$ field directions. (c) The temperature dependence of the lower critical field is shown for both directions and the red dotted line is a fit using the GL equation.}
\end{figure*}

The field dependence of magnetization was used to estimate the lower critical field $(H_{c1})$. \figref{fig2}(a), and \figref{fig2}(b) show the different M(\textit{H}) curves measured under different temperatures with fields along $H || ab$ and $H || c$ directions, respectively. The criteria for determining $H_{c1}$ was based on the point at which the M(\textit{H}) curve drifts from the linear Meissner line. The obtained temperature variation of $H_{c1}$ for both field directions is shown in \figref{fig2}(c). Using the Ginzburg-Landau (GL) equation \equref{eqn1} \cite{hc1_formula} to fit, the estimated $H_{c1}$(0) values are: $H_{c1}^{|| ab}$(0) = 15.2(1) Oe, and $H_{c1}^{|| c}$(0) = 11.3(1) Oe.
\begin{equation}
    H_{c1}(T)= H_{c1}(0)\left[1-\left(\frac{T}{T_{c}}\right)^2\right].
\label{eqn1}
\end{equation}

\subsection{Transport measurement}
The temperature dependence of the zero-field resistivity above 10 K can be described by using the Bloch-Grüneisen (BG) model which defines resistivity as \cite{bg_model}
\begin{equation*}
    \rho(T) = \rho_0 + \rho_{BG}(T),
\end{equation*}
where $\rho_0$ is the $T$-independent residual resistivity, and $\rho_{BG}(T)$ can be written as:
\begin{equation}
    \rho_{BG}(T)= 4C\left(\frac{T}{\theta_{D}^R}\right)^r\int_{0}^{\theta_{D}^R/{T}}\frac{x^{r}}{(e^{x}-1)(1-e^{-x})}\,dx.
\label{eqn2}
\end{equation}
Here, $C$ is a coupling constant, and $\theta_D^R$ denotes the Debye temperature. The data is well fitted for $r$ = 3, indicating dominant $s$-$d$ interband scattering. From the fitting, the extracted values are $\rho_0$ = 57.24(15) $\mu\Omega$ cm, $C$ = 0.15 m$\Omega$ cm and $\theta_D^R$ = 176.4(7) K.

\begin{figure*}
    \centering
    \includegraphics[width=0.8\textwidth]{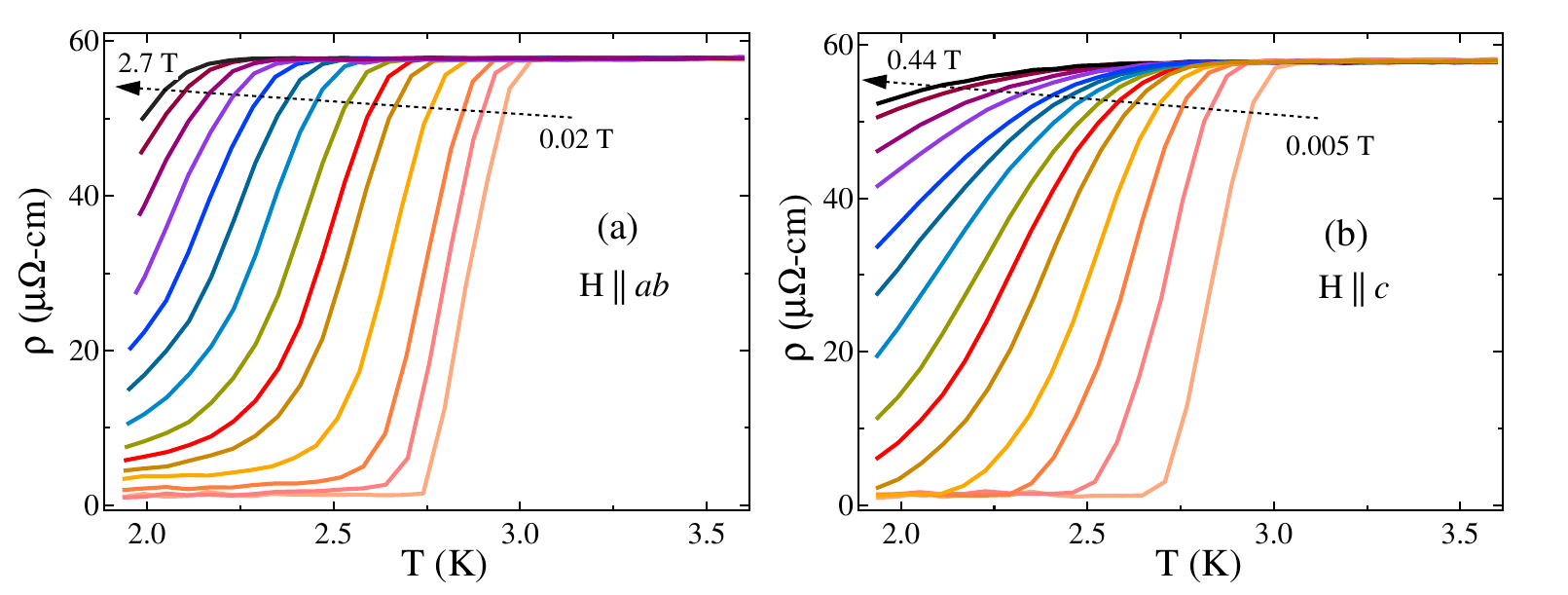}
    \caption{\label{fig3}temperature dependence of resistivity for (PbS)$_{1.13}$TaS$_2$ crystal in the superconducting state under different fields, applied along (a) $H || ab$, and $H || c$ directions.}
\end{figure*}

The resistivity was measured with different magnetic fields to determine the upper critical field $H_{c2}$(0) for bulk (PbS)$_{1.13}$TaS$_2$ crystal. \figref{fig3} presents the different temperature-dependent resistivity curves for two field orientations. A two-band model was followed to explain $T$ dependence of $H_{c2}$, which can be expressed as \cite{2gap_model,TaSeS}:
\begin{multline}
 \ln{\frac{T_c}{T}}=\frac{1}{2}\left[U(s)+U(\eta s)+\frac{\lambda_0}{\omega}\right]\\-\left\{\frac{1}{4}\left[U(s)-U(\eta s)-\frac{\lambda_-}{\omega}\right]^2+\frac{\lambda_{12}\lambda_{21}}{\omega^2}\right\}^{1/2},\\\textrm{where}\hspace{0.5cm} s = \frac{H_{c2}D_1}{2\phi_0T}; \hspace{0.1cm}U(s) = \psi\left({s+\frac{1}{2}}\right)-\psi\left(\frac{1}{2}\right).
 \label{eqn6}
\end{multline}
Here $\psi(x)$ is the digamma function, and $s$ is the reduced critical magnetic field. The coupling constants are defined as $\lambda_- = (\lambda_{11}-\lambda_{22})$, $\lambda_0=(\lambda_-^2+4\lambda_{12}\lambda_{21})^{1/2}$, and $\omega=\lambda_{11}\lambda_{22}-\lambda_{12}\lambda_{21}$. $\lambda_{11}$ and $\lambda_{22}$ are the intraband coupling constants, while $\lambda_{12}$ and $\lambda_{21}$ are the interband coupling constants. $\eta = D_2/D_1$ is the ratio of diffusivity of each band and $\phi_0$ is the flux quantum. The deduced $H_{c2}$(0) values are: $H_{c2}^{|| ab}$(0) = 7.37(8) T, and $H_{c2}^{||c}$(0) = 0.98(3) T.

By using $H_{c1}$(0) and $H_{c2}$(0) values, the superconducting characteristic length parameters, the penetration length $\lambda$(0), and the Ginzburg coherence length $\xi$(0) can be calculated. For layered superconductors, the coherence length can be obtained using equations \equref{eqn3}, based on an anisotropic GL theory. the resulting values of $\xi_{c}$ and $\xi_{ab}$ are 2.44(6) nm and 18.34(28) nm, respectively.
\begin{equation}
    H_{c2}^{|| c} = \phi_0/2\pi \xi_{ab}^2;\hspace{0.5cm} H_{c2}^{|| ab} = \phi_0/2\pi \xi_{c} \xi_{ab}
    \label{eqn3}
\end{equation}
 The penetration depth was estimated by the following equations:
\begin{equation}
    H_{c1}^{||c}(0)=\frac{\phi_0}{4\pi\lambda_{ab}^2(0)}\left[\ln{\left(\frac{\lambda_{ab}}{\xi_{ab}}\right)}\right].
\end{equation}
\begin{equation}
    H_{c1}^{|| ab}(0)=\frac{\phi_0}{4\pi\lambda_{ab}(0)\lambda_{c}(0)}\left[\ln{\left(\frac{\lambda_{ab}}{\xi_{c}}\right)}\right].
\end{equation}
Here, $\phi_0 = h/2e$ is the flux quantum. The obtained values of  $\lambda_{c}$ and $\lambda_{ab}$ are 843.5(6) nm, and 733.4(6) nm, respectively. The GL parameter ($\kappa$) was determined by taking the ratio of both length scales, yielding $\kappa_{c}$ = $\lambda_{ab}/\xi_{ab}$ = 40 $>$ 1/$\sqrt{2}$. This indicates type-II superconductivity in bulk (PbS)$_{1.13}$TaS$_2$.

Further, to explain the angular dependence of $H_{c2}$, two models three-dimensional (3D) anisotropic GL and a two-dimensional (2D) Tinkham model were applied. According to the 3D model, for the 3D superconductors, the $H_{c2}(\theta)$ can be represented as \cite{3dmodel}: 
\begin{equation}
    \left(\frac{H_{c2}(\theta)  \sin{\theta}}{H_{c2}^\perp}\right)^{2} +  \left(\frac{H_{c2}(\theta)  \cos{\theta}}{H_{c2}^\parallel}\right)^{2} = 1
    \label{eq6}
\end{equation}
In this model, $H_{c2}$ varies smoothly with different field orientations. Meanwhile, for the 2D thin-film superconductors, where thickness is much smaller than the coherence length, Tinkham \cite{2dmodel} proposed the following equation: 
\begin{equation}
    \left(\frac{H_{c2}(\theta)  \sin{\theta}}{H_{c2}^\perp}\right)^{2} + \left|\frac{H_{c2}(\theta) \cos{\theta}}{H_{c2}^\parallel}\right| = 1
    \label{eq7}
\end{equation}
Our $H_{c2}(\theta)$ data is well defined by 2D Tinkham model, indicating 2D superconductivity in (PbS)$_{1.13}$TaS$_2$.

\subsection{Specific heat measurement}
\begin{figure*}
    \centering
    \includegraphics[width=0.8\textwidth]{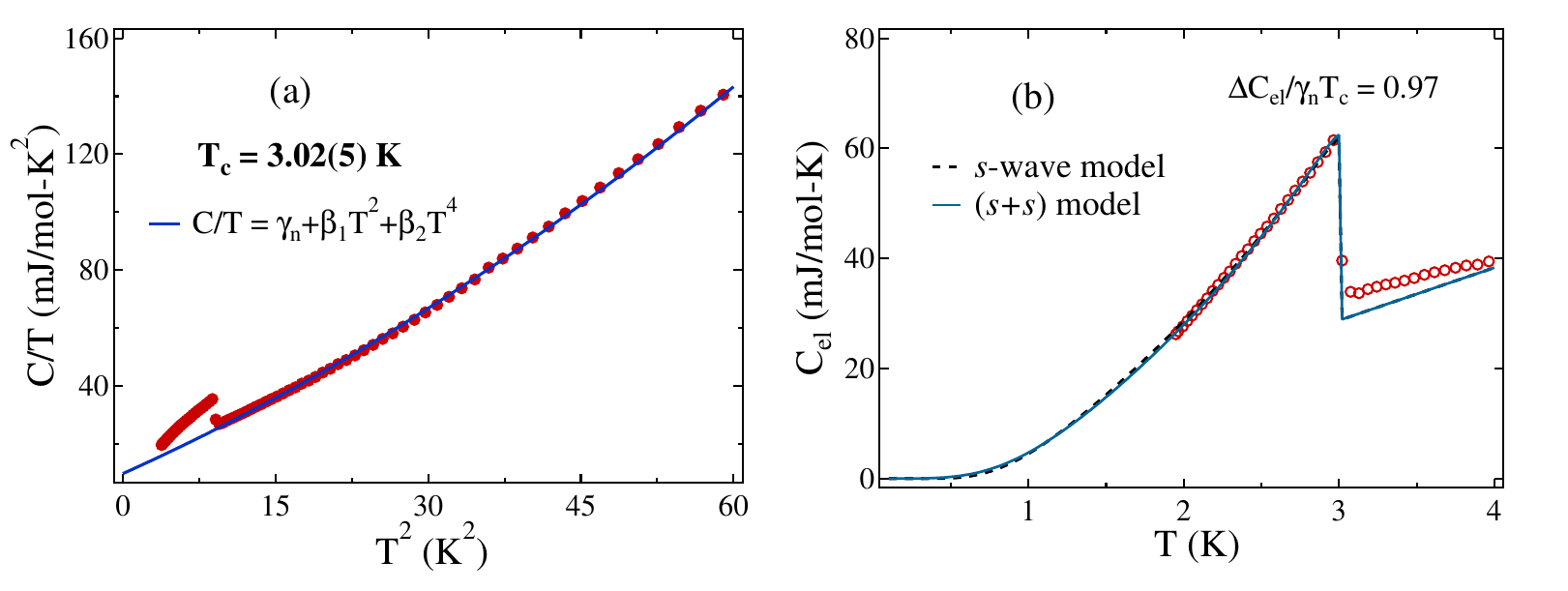}
    \caption{\label{fig4}(a) The crystal structure of misfit compound (PbS)$_{1.13}$TaS$_2$ formed by alternate stacking of PbS and TaS$_2$ layers. (b) XRD pattern of a single crystal of (PbS)$_{1.13}$TaS$_2$ and the inset shows a microscopic image of the grown crystal.}
\end{figure*}

The zero field specific heat measurement was carried out to ensure the bulk superconducting transition in misfit compound (PbS)$_{1.13}$TaS$_2$. \figref{fig4} presents a clear discontinuity in the measured specific heat, marking the transition into the superconducting state at $T_c$ = 3.02(5) K, consistent with resistivity and magnetization $T_c$ values. The general expression of the total specific heat, incorporating contributions from both phononic and electronic components, can be defined as follows: $\frac{C}{T}$ = $\gamma_{n}$+$\beta_1 T^2$+$\beta_2 T^4$. In this expression, $\gamma_{n}$ denotes the Sommerfeld coefficient, $\beta_1$ is the Debye constant, associated with phononic contribution and $\beta_2$ represents the anharmonic contribution. By fitting the normal-state specific heat data to this expression, the obtained values are $\gamma_n$ = 10.21(53) mJ mol$^{-1}$ K$^{-2}$, $\beta_1$ = 1.53(3) mJ mol$^{-1}$ K$^{-4}$, and $\beta_2$ = 1.16(5) $\mu$J mol$^{-1}$ K$^{-6}$. 
The Debye constant $\beta_1$ can be related to the Debye temperature ($\theta_D$) through the following relation  $ \theta_D^{SH} = \left(\frac{12\pi^4RN}{5\beta}\right)^{1/3}$, where $R$ is the universal gas constant and $N$ is the number of atoms per unit cell. Considering $N$ = 5, the Debye temperature is calculated to be 185(1) K, similar to the value obtained from resistivity data. Using $\gamma_{n}$ value, the density of states (DOS) at the Fermi level is calculated from the relation $\gamma_n$ = $\left(\frac{\pi^2k_B^2}{3}\right) D(E_F)$ to be 4.34(22) $eV^{-1}fu^{-1}$. The electron-phonon coupling constant can be estimated by McMillan's equation \equref{eqn5}\cite{mcmillan}, proposed for phonon-mediated, single-band superconductors as follows:
\begin{equation}
 \lambda_{e-ph} = \frac{1.04 + \mu^{*}\ln{(\theta_{D}/1.45T_{c})}}{(1 - 0.62\mu^{*})\ln{(\theta_{D}/1.45T_{c}}) - 1.04 } ~,
 \label{eqn5}
\end{equation}
where $\mu^*$ is the repulsive screened Coulomb potential. By considering $\mu^*$ = 0.13, the obtained value of $\lambda_{e-ph}$ is 0.63(6), suggests (PbS)$_{1.13}$TaS$_2$ is a weakly coupled superconductor.

The superconducting gap parameters of (PbS)$_{1.13}$TaS$_2$ were analyzed by determining the electronic contribution of the specific heat ($C_{el} (T)$) below $T_c$. The $C_{el}$ was estimated by subtracting the phononic contribution from the total specific heat, measured at zero field, which is shown in \figref{fig4}(b). The specific heat jump $\Delta C_{el}/\gamma_n T_c$ was calculated to be 0.97, much smaller than the BCS value of 1.43. Similarly low values have also been reported for other MLCs \cite{SnNb2Se5,SnTi2Se5,SnNbSe3}. 

The temperature dependence of $C_{el}$ was fitted with the \textit{s}-wave model, which defines the entropy $S$ for a fully-gapped BCS superconductor as:
\begin{equation}
    S = -\frac{6 \gamma}{\pi^2}\left(\frac{\Delta(0)}{k_B}\right)\int_{0}^{\infty}{[f\ln{f} + (1 - f) \ln{(1-f)}]}\,dy
    \label{eqn10}.
\end{equation}
Here $f(\xi)=[\exp{(E(\xi)/k_{B}T)} + 1]^{-1}$ is the Fermi function, $E(\xi) = \sqrt{\xi^2 + \Delta^2(t)}$ is the quasiparticles's excitation energy, $t = T/T_c$, and $\Delta(t)=\tanh{[1.82(1.018(1/t) -1))^{0.51}]}$ defines the $T$-dependence of the superconducting gap. By differentiating the entropy, $C_{el}(T)$ can be determined by $C_{el}=t\left(\frac{dS}{dt}\right)$. In \figref{fig4}(b), the black dashed line represents the $s$-wave model fitting, showing slight deviation at lower temperatures. We further applied the two-gap ($s+s$) model to elucidate the superconducting gap symmetry \cite{2gap_SH_1,2gap_SH_2}. The two-gap model accounts for the presence of two distinct energy gaps originating from separate electronic bands. In this model, the total specific heat can be defined as the sum of contributions from two bands, each characterized by its own gap parameter ($\Delta_1$ and $\Delta_2$) with relative weights $f$. The fit using the two-gap model is depicted by a blue solid line in \figref{fig4}(b), showing a better agreement with the data. The estimated gap values are $\Delta_1$ = 0.43(4) meV and $\Delta_2$ = 0.20(8) meV, with a weight fraction $f$ = 0.87 and $\gamma$ = 9.58(1) mJ mol$^{-1}$ K$^{-2}$. Nevertheless, low-temperature measurements ($T < 0.2T_c$) are necessary to determine the precise nature of the superconducting gap.

\begin{table}
\caption{Summary of physical parameters for bulk misfit compound (PbS)$_{1.13}$TaS$_2$.}
\label{tbl1}
\setlength{\tabcolsep}{10pt}
\begin{center}
\begin{tabular}{lcl}\hline
Parameters& Unit &(PbS)$_{1.13}$TaS$_2$ \\
\hline
$T_{c}$&K  & 3.05(5)\\
RRR & & 9.3\\
 $H_{c2}^{|| ab}$ & T & 7.37(8) \\
  $H_{c2}^{|| c}$& T & 0.98(3) \\
   $H_{P}$& T & 5.67 \\
   $\Gamma$& & 7.5\\
   $\xi_{ab}(0)$& nm &18.34(28) \\
   $\xi_{c}(0)$& nm & 2.44(6)\\
   $\lambda_{ab}(0)$& nm &733.4(6) \\
   $\lambda_{c}(0)$& nm &843.5(6) \\
   $\gamma_n$& mJ/mol-K$^2$ &10.21(53) \\
   $\beta_1$& mJ/mol-K$^4$ &1.53(3) \\
   $\beta_2$& $\mu$J/mol-K$^6$ &1.16(5) \\
   $\theta_D^{SH}$&K & 185(1)\\
   $D(E_F)$& $eV^{-1}fu^{-1}$&4.34(22)\\
   $\lambda_{e-ph}$&  &0.63(6) \\
   $\Delta C_{el}/\gamma_n T_c$& & 0.97\\
   $n$(10 K)& 10$^{22}$cm$^{-3}$ &1.74(8) \\
\hline
\end{tabular}
\end{center}
\end{table}


\begin{thebibliography}{References}

\bibitem{vdW_H1}K. S. Novoselov, A. Mishchenko, A. Carvalho, and A. H. Castro Neto, 2D materials and van der Waals heterostructures, \href{ http://dx.doi.org/10.1126/science.aac9439}{Science \textbf{353}, aac9439 (2016).}

\bibitem{vdW_H2}S. Kezilebieke, M.N. Huda, V. Vaňo, M. Aapro, S.C. Ganguli,  O.J. Silveira, S. Głodzik, A.S. Foster, T. Ojanen, and P. Liljeroth, Topological superconductivity in a van der Waals heterostructure, \href{https://doi.org/10.1038/s41586-020-2989-y}{Nature \textbf{588}, 424 (2020).}

\bibitem{vdW_H3}S. P. Chiu, C. C. Tsuei, S. S. Yeh, F. C. Zhang, S. Kirchner, and J.J. Lin, Observation of triplet superconductivity in CoSi$_2$/TiSi$_2$ heterostructures, \href{https://doi.org/10.1126/sciadv.abg6569}{Sci. Adv. \textbf{7}, eabg6569 (2021).}

\bibitem{misfit1}N. Ng and T.M. McQueen, Misfit layered compounds: Unique, tunable heterostructured materials with untapped properties, \href{https://doi.org/10.1063/5.0101429}{APL Materials \textbf{10}, 100901 (2022)}

\bibitem{misfit2}O. Dolotko, I.Z. Hlova, A.K. Pathak, Y. Mudryk, V.K. Pecharsky, P. Singh, D.D. Johnson, B.W. Boote, J. Li, E.A. Smith, S.L. Carnahan, A.J. Rossini, L. Zhou, E.M. Eastman, and V.P. Balema, Unprecedented generation of 3D heterostructures by mechanochemical disassembly and re-ordering of incommensurate metal chalcogenides, \href{https://doi.org/10.1038/s41467-020-16672-0}{Nature Communications \textbf{11}, 3005 (2020).}

\bibitem{misfit3}R. T. Leriche, A. Palacio-Morales, M. Campetella, C. Tresca, S. Sasaki, C. Brun, F. Debontridder, P. David, I. Arfaoui, O. Šofranko, T. Samuely, G. Kremer, C. Monney, T. Jaouen, L. Cario, M. Calandra, and T. Cren, Misfit Layer Compounds: A Platform for Heavily Doped 2D Transition Metal Dichalcogenides, \href{https://doi.org/10.1002/adfm.202007706}{Adv. Funct. Mater. \textbf{31}, 2007706 (2021).}

\bibitem{SC_misfit1}N. Ng, and T. M. McQueen, Misfit layered compounds: Unique, tunable heterostructured materials with untapped properties, \href{https://doi.org/10.1063/5.0101429}{APL Mater. \textbf{10}, 100901 (2022).}

\bibitem{SC_misfit2}A. Devarakonda, H. Inoue, S. Fang, C. Ozsoy-Keskinbora, T. Suzuki, M. Kriener, L. Fu, E. Kaxiras, D. C. Bell and J. G. Checkelsky, Clean 2D superconductivity in a bulk van der Waals superlattice, \href{https://doi.org/10.1126/science.aaz6643}{Science 370, 231 (2020).}

\bibitem{SOC_TMD1}E. Cappelluti, R. Roldán, J. A. Silva-Guillén, P. Ordejón, and F. Guinea, Tight-binding model and direct-gap/indirect-gap transition in single-layer and multilayer MoS$_2$, \href{https://doi.org/10.1103/PhysRevB.88.075409}{Phys. Rev. B 88, 075409 (2013).}

\bibitem{SOC_TMD2}D. Xiao, G.-B. Liu, W. Feng, X. Xu, and W. Yao, Coupled Spin and Valley Physics in Monolayers of MoS$_2$ and Other Group-VI Dichalcogenides, \href{https://doi.org/10.1103/PhysRevLett.108.196802}{Phys. Rev. Lett. 108, 196802 (2012).}

\bibitem{ising_misfit1}X. Sun, Z. Deng, Y. Yang, S. Yu, Y. Huang, Y. Lu, Q. Tao, D.-W. Shen, W.-Y. He, C. Xi, L. Pi, K. Watanabe, T. Taniguchi, Z.-A. Xu, and Y. Zheng, Tunable Mirror-Symmetric Type-III Ising Superconductivity in Atomically-Thin Natural Van der Waals Heterostructures, \href{https://doi.org/10.1002/adma.202411655}{Adv. Mater., 2411655 (2024).}

\bibitem{ising_misfit2}T. Samuely, D. Wickramaratne, M. Gmitra, T. Jaouen, O. Šofranko, D. Volavka, M. Kuzmiak, J. Haniš, P. Szabó, C. Monney, G. Kremer, P. Le Fèvre, F. Bertran, T. Cren, S. Sasaki, L. Cario, M. Calandra, Igor I. Mazin, and P. Samuely, Protection of Ising spin-orbit coupling in bulk misfit superconductors, \href{https://doi.org/10.1103/PhysRevB.108.L220501}{Phys. Rev. B \textbf{108}, L220501 (2023).}

\bibitem{SOC_misfit1}H. Zhong, H. Zhang, H. Zhang, T. Bao, K. Zhang, S. Xu, L. Luo, A. Rousuli, W. Yao, and J.D. Denlinger, Y. Huang, Y. Wu, Y. Xu, W. Duan, and S. Zhou, Revealing the two-dimensional electronic structure and anisotropic superconductivity in a natural van der Waals superlattice (PbSe)$_{1.14}$NbSe$_2$, \href{https://doi.org/10.1103/PhysRevMaterials.7.L041801}{Phys. Rev. Materials \textbf{7}, L041801 (2023).}

\bibitem{SOC_misfit2}Sajilesh K. P., R. A. Gofman, Y. Nitzav, A. Almoalem, I. Mangel, T. Shiroka, N. C. Plumb, C. Bigi, F. Bertran, J. Sánchez-Barriga, and A. Kanigel, Ising superconductivity in the bulk incommensurate layered material (PbS)$_{1.13}$(TaS$_2$), \href{https://doi.org/10.48550/arXiv.2411.07624}{	arXiv:2411.07624 (2024).}

\bibitem{SOC_misfit3}L. Fang, J. Im, W. DeGottardi, Y. Jia, A. Glatz, K. A. Matveev, W.-K. Kwok, G. W. Crabtree, and M. G. Kanatzidis, Large spin-orbit coupling and helical spin textures in 2D heterostructure [Pb$_2$BiS$_3$][AuTe$_2$], \href{https://doi.org/10.1038/srep35313}{Sci Rep \textbf{6}, 35313 (2016)}.

\bibitem{LaNSe3}M. Shan, S. Li, Y. Yang, D. Zhao, J. Li, L. Nie, Z. Wu, Y. Zhou, L. Zheng, B. Kang, T. Wu, and X. Chen, Anisotropic Spin Fluctuations Induced by Spin-Orbit Coupling in a Misfit Layer Compound (LaSe)$_{1.14}$(NbSe$_2$), \href{https://doi.org/10.1002/advs.202403824}{Adv. Sci. \textbf{11}(40), 2403824 (2024).}

\bibitem{SnNbS3}S. Li, X. Wang, Z. Yang, L. Zhang, S.L. Teo, M. Lin, R. He, N. Wang, P. Song, W. Tian, X.J. Loh, Q. Zhu, B. Sun, and X. R. Wang, Giant Third-Order Nonlinear Hall Effect in Misfit Layer Compound (SnS)$_{1.17}$(NbS$_2$)$_3$, \href{ https://pubs.acs.org/doi/10.1021/acsami.3c18319}{ ACS Appl. Mater. Interfaces \textbf{16}, 11043 (2024).}

\bibitem{PbTaSe2}Y. M. Itahashi, T. Ideue, S. Hoshino, C. Goto, H. Namiki, T. Sasagawa, and Y. Iwasa, Giant second harmonic transport under time-reversal symmetry in a trigonal superconductor, \href{https://doi.org/10.1038/s41467-022-29314-4}{Nat. Commun. \textbf{13}, 1659 (2022).}

\bibitem{SOC_TM1}C. Tan, M.X. Deng, G. Zheng, F. Xiang, S. Albarakati, M. Algarni, L. Farrar, S. Alzahrani, J. Partridge, J. B. Yi, A. R. Hamilton, R.Q. Wang, and L. Wang, Spin-Momentum Locking Induced Anisotropic Magnetoresistance in Monolayer WTe$_2$, \href{https://doi.org/10.1021/acs.nanolett.1c02329}{Nano Lett. \textbf{21}, 9005 (2021).}

\bibitem{SOC_TM2}K. Premasiri, and Xuan P A Gao, Tuning spin–orbit coupling in 2D materials for spintronics: a topical review, \href{https://doi.org/10.1088/1361-648X/ab04c7}{J. Phys.: Condens. Matter \textbf{31}, 193001 (2019).}

\bibitem{SOC_TM3}X. Qian, J. Liu, L. Fu, and J. Li, Quantum spin Hall effect in two-dimensional transition metal dichalcogenides, \href{https://doi.org/10.1126/science.1256815}{Science \textbf{346}(6215), 1344 (2014)}

\bibitem{transport_TM1}X. C. Yang, X. Luo, J. J. Gao, Z. Z. Jiang, W. Wang, T. Y. Wang, J. G. Si, C. Y. Xi, W. H. Song, and Y. P. Sun, Planar Hall effect in the quasi-one-dimensional topological superconductor TaSe$_3$,  \href{https://doi.org/10.1103/PhysRevB.104.155106}{Phys. Rev. B \textbf{104}, 155106 (2021).}

\bibitem{transport_TM2}H. Li, H.-W. Wang, H. He, J. Wang, and S.-Q. Shen, Giant anisotropic magnetoresistance and planar Hall effect in the Dirac semimetal Cd$_3$As$_2$, \href{https://doi.org/10.1103/PhysRevB.97.201110}{Phys. Rev. B \textbf{97}, 201110(R) (2018).}

\bibitem{TaS2thin_SC}E. Navarro-Moratalla, J.O. Island, S. Mañas-Valero, E. Pinilla-Cienfuegos, A. Castellanos-Gomez, J. Quereda, G. Rubio-Bollinger, L. Chirolli, J.A. Silva-Guillén, N. Agraït, G. A. Steele, F. Guinea, H. S. J. van der Zant, and E. Coronado, Enhanced superconductivity in atomically thin TaS$_2$, \href{https://doi.org/10.1038/ncomms11043}{Nat. Commun. \textbf{7}, 11043 (2016).}

\bibitem{TaS2_ising}S. C. de la Barrera, M. R. Sinko, D. P. Gopalan, N. Sivadas, K.L. Seyler, K. Watanabe, T. Taniguchi, A. W. Tsen, X. Xu, D. Xiao, and B. M. Hunt, Tuning Ising superconductivity with layer and spin-orbit coupling in two-dimensional transition-metal dichalcogenides, \href{https://doi.org/10.1038/s41467-018-03888-4}{Nat Commun \textbf{9}, 1427 (2018).}

\bibitem{DFT_PbTS3}E. Kabliman, P. Blaha, and K. Schwarz, Ab initio study of stabilization of the misfit layer compound (PbS)$_{1.14}$TaS$_2$, \href{https://doi.org/10.1103/PhysRevB.82.125308}{Phys. Rev. B \textbf{82}, 125308 (2010).}

\bibitem{TaS2bulk_SC}S. Nagata, T. Aochi, T. Abe, S. Ebisu, T. Hagino, Y. Seki, and K. Tsutsumi, Superconductivity in the layered compound 2H-TaS$_2$, \href{https://doi.org/10.1016/0022-3697(92)90242-6}{J. Phys. Chem. Solids 53, 1259 (1992).}

\bibitem{monolayer_tas2}Y. Yang, S. Fang, V. Fatemi, J. Ruhman, E. Navarro-Moratalla, K. Watanabe, T. Taniguchi, E. Kaxiras, and P. Jarillo-Herrero, Enhanced superconductivity upon weakening of charge density wave transport in 2\textit{H}-TaS$_2$ in the two-dimensional limit, \href{https://doi.org/10.1103/PhysRevB.98.035203}{Phys. Rev. B \textbf{98}, 035203 (2018).}

\bibitem{gap_TaS2_1}C.S. Lian, C. Heil, X. Liu, C. Si, F. Giustino, and W. Duan, Intrinsic and doping-enhanced superconductivity in monolayer 1\textit{H}-TaS$_2$: Critical role of charge ordering and spin-orbit coupling, \href{https://doi.org/10.1103/PhysRevB.105.L180505}{Phys. Rev. B 105, L180505 (2022).}

\bibitem{gap_TaS2_2}A. Ribak, R. Majlin Skiff, M. Mograbi, P. K. Rout, M. H. Fischer, J. Ruhman, K. Chashka, Y. Dagan, and A. Kanigel, Chiral superconductivity in the alternate stacking compound 4Hb-TaS$_2$, \href{https://www.science.org/doi/10.1126/sciadv.aax9480}{Sci. Adv. 6, eaax9480 (2020).}

\bibitem{2gap1}A. Gurevich, S. Patnaik, V. Braccini, K. H. Kim, C. Mielke, X. Song, L. D. Cooley, S. D. Bu, D. M. Kim, J. H. Choi, L. J. Belenky, J. Giencke, M. K. Lee, W. Tian, X. Q. Pan, A. Siri, E. E. Hellstrom, C. B. Eom, and D. C. Larbalestier, Very high upper critical fields in MgB$_2$ produced by selective tuning of impurity scattering, \href{https://doi.org/10.1088/0953-2048/17/2/008}{Supercond. Sci. Technol. \textbf{17}, 278 (2003).}

\bibitem{2gap2}F. Hunte, J. Jaroszynski, A. Gurevich, D. C. Larbalestier, R. Jin, A. S. Sefat, M. A. McGuire, B. C. Sales, D. K. Christen, and D. Mandrus, Two-band superconductivity in LaFeAsO$_{0.89}$F$_{0.11}$ at very high magnetic fields, \href{https://doi.org/10.1038/nature07058}{Nature \textbf{453}, 903 (2008).}

\bibitem{2gap3}H. Bai, L. Qiao, M. Li, J. Ma, X. Yang, Y. Li, Q. Tao, and Z.-A. Xu, Multi-band Superconductivity in a misfit layered compound (SnSe)$_{1.16}$(NbSe$_2$)$_2$, \href{https://doi.org/10.1088/2053-1591/ab60a8}{Mater. Res. Express 7, 016002 (2020).}

\bibitem{PbNbSe3}H. Zhong, H. Zhang, H. Zhang, T. Bao, K. Zhang, S. Xu, L. Luo, A. Rousuli, W. Yao, J. D. Denlinger, Y. Huang, Y. Wu, Y. Xu, W. Duan, and S. Zhou, Revealing the two-dimensional electronic structure and anisotropic superconductivity in a natural van der Waals superlattice (PbSe)$_{1.14}$NbSe$_2$, \href{https://doi.org/10.1103/PhysRevMaterials.7.L041801}{Phys. Rev. Materials \textbf{7}, L041801 (2023).}

\bibitem{3dmodel}M. Tinkham, Introduction to Superconductivity (Courier Corporation, 2004).

\bibitem{2dmodel}M. Tinkham, Effect of fluxoid quantization on transitions of superconducting films, \href{https://link.aps.org/doi/10.1103/PhysRev.129.2413}{Phys. Rev. \textbf{129}, 2413 (1963).}

\bibitem{2d_misfit1}P. Samuely, P. Szabó, J. Kačmarčík, A. Meerschaut, L. Cario, A. G. M. Jansen, T. Cren, M. Kuzmiak, O. Šofranko, and T. Samuely, Extreme in-plane upper critical magnetic fields of heavily doped quasi-two-dimensional transition metal dichalcogenides, \href{https://doi.org/10.1103/PhysRevB.104.224507}{Phys. Rev. B \textbf{104}, 224507 (2021).} 

\bibitem{2d_misfit2}S. Matsuzawa, S. Pyon, and T. Tamegai, Characterizations of Anisotropic Superconductivity in (BiSe)$_{1+\delta}$NbSe$_2$, \href{https://doi.org/10.1088/1742-6596/2323/1/012008}{J. Phys.: Conf. Ser. \textbf{2323}, 012008 (2022).}

\bibitem{WSM_PHE}S. Nandy, G. Sharma, A. Taraphder, and S. Tewari, Chiral anomaly as the origin of the planar Hall effect in Weyl semimetals, \href{https://doi.org/10.1103/PhysRevLett.119.176804}{Phys. Rev. Lett. \textbf{119}, 176804 (2017).}

\bibitem{WSM1}J. Yan, X. Luo, J. J. Gao, H. Y. Lv, C. Y. Xi, Y. Sun, W. J. Lu, P. Tong, Z. G. Sheng, X.B. Zhu, and W.H. Song, The giant planar Hall effect and anisotropic magnetoresistance in Dirac node arcs semimetal PtSn$_4$, \href{https://doi.org/10.1088/1361-648X/ab851f}{J. Phys.: Condens. Matter \textbf{32}, 315702 (2020).}

\bibitem{WSM2}D. D. Liang, Y. J. Wang, W. L. Zhen, J. Yang, S. R. Weng, X. Yan, Y. Y. Han, W. Tong, W. K. Zhu, L. Pi, and C. J. Zhang, Origin of planar Hall effect in type-II Weyl semimetal MoTe$_2$, \href{https://doi.org/10.1063/1.5094231}{AIP Advances \textbf{9}, 055015 (2019).}

\bibitem{PHE_formula} A. A. Burkov, Giant planar Hall effect in topological metals, \href{https://doi.org/10.1103/PhysRevB.96.041110}{Phys. Rev. B \textbf{96}, 041110(R) (2017).}

\bibitem{fit_eq}X. Ma, M. Huang, S. Wang. P. Liu, Y. Zhang, Y. Lu, and B. Xiang, Anisotropic Magnetoresistance and Planar Hall Effect in Layered Room-Temperature Ferromagnet Cr$_{1.2}$Te$_2$, \href{https://doi.org/10.1021/acsaelm.3c00312}{ACS Appl. Electron. Mater. 5, 2838 (2023).}

\bibitem{butterfly_amr1}K. Mukherjee, S. D. Das, N. Mohapatra, K. K. Iyer, and E. V. Sampathkumaran, Anomalous butterfly-shaped magnetoresistance loops in the alloy Tb$_4$LuSi$_3$, \href{https://doi.org/10.1103/PhysRevB.81.184434}{Phys. Rev. B \textbf{81}, 184434 (2010).}

\bibitem{butterfly_amr2}V. P. Jovanović, L. Fruchter, Z. Z. Li, and H. Raffy, Anisotropy of the in-plane angular magnetoresistance of electron-doped Sr$_{1-x}$La$_x$CuO$_2$ thin films, \href{https://doi.org/10.1103/PhysRevB.81.134520}{Phys. Rev. B \textbf{81}, 134520 (2010).}

\bibitem{butterfly_amr3}P. Li, C. Jin, E. Y. Jiang, and H. L. Bai, Origin of the twofold and fourfold symmetric anisotropic magnetoresistance in epitaxial Fe$_3$O$_4$ films, \href{https://doi.org/10.1063/1.3499696}{J. Appl. Phys. \textbf{108}, 093921 (2010).}

\bibitem{butterfly_amr4}Y. Dai, Y. W. Zhao, L. Ma, M. Tang, X. P. Qiu, Y. Liu, Z. Yuan, and S. M. Zhou, Fourfold Anisotropic Magnetoresistance of L1$_0$ FePt Due to Relaxation Time Anisotropy, \href{https://doi.org/10.1103/PhysRevLett.128.247202}{Phys. Rev. Lett. \textbf{128}, 247202 (2022).}

\bibitem{butterfly_amr5}Y. Yahagi, D. Miura, and A. Sakuma, Theoretical Study on Four-fold Symmetric Anisotropic Magnetoresistance Effect in Cubic Single-crystal Ferromagnetic Model, \href{https://doi.org/10.7566/JPSJ.89.044714}{J. Phys. Soc. Jpn. \textbf{89}, 044714 (2020).}


\bibitem{ZrSiS}M. N. Ali, L. M. Schoop, C. Garg, J. M. Lippmann, E. Lara, B. Lotsch, and S. S. P. Parkin, Butterfly magnetoresistance, quasi-2D Dirac Fermi surface and topological phase transition in ZrSiS, \href{https://doi.org/10.1126/sciadv.1601742}{Sci. Adv. \textbf{2}(12), e1601742 (2016).}

\bibitem{ZrTe5}G. Zheng, X. Zhu, Y. Liu, J. Lu, W. Ning, H. Zhang, W. Gao, Y. Han, J. Yang, H. Du, K. Yang, Y. Zhang, and M. Tian, Field-induced topological phase transition from a three-dimensional Weyl semimetal to a two-dimensional massive Dirac metal in ZrTe$_5$, \href{https://doi.org/10.1103/PhysRevB.96.121401}{Phys. Rev. B \textbf{96}, 121401(R) (2017).}

\bibitem{MoAs2}R. Singha, A. Pariari, G. K. Gupta, T. Das, and P. Mandal, Probing the Fermi surface and magnetotransport properties of MoAs$_2$, \href{https://doi.org/10.1103/PhysRevB.97.155120}{Phys. Rev. B \textbf{97}, 155120 (2018).}

\bibitem{PHE1}M. Mandal, C. Patra, A. Kataria, S. Paul, S. Saha, and R. P. Singh, Superconductivity in doped Weyl semimetal Mo$_{0.9}$Ir$_{0.1}$Te$_2$ with broken inversion symmetry, \href{https://iopscience.iop.org/article/10.1088/1361-6668/ac3b38}{Supercond. Sci. Technol. \textbf{35}, 025011, (2022).}

\bibitem{PHE2}J. Meng, H. Xue, M. Liu, W. Jiang, Z. Zhang, J. Ling, L. He, R. Dou, C. Xiong, and J. Nie, Planar Hall effect induced by anisotropic orbital magnetoresistance in type-II Dirac semimetal PdTe$_2$, \href{https://doi.org/10.1088/1361-648X/ab4464}{J. Phys.: Condens. Matter \textbf{32}, 015702 (2019).}

\bibitem{MR1}W. Ren, A. Wang, D. Graf, Y. Liu, Z. Zhang, W.-G. Yin, and  C. Petrovic, Absence of Dirac states in BaZnBi$_2$ induced by spin-orbit coupling, \href{https://doi.org/10.1103/PhysRevB.97.035147}{Phys. Rev. B \textbf{97}, 035147 (2018).}  

\bibitem{MR2}S.-M. Huang, S.-H. Yu, and M. Chou, The linear magnetoresistance from surface state of the Sb$_2$SeTe$_2$ topological insulator, \href{https://doi.org/10.1063/1.4954290}{J. Appl. Phys. \textbf{119}, 245110 (2016).}  

\bibitem{MR3}J. Jiang, F. Tang, X. C. Pan, H. M. Liu, X. H. Niu, Y. X. Wang, D. F. Xu, H. F. Yang, B. P. Xie, F. Q. Song, P. Dudin, T. K. Kim, M. Hoesch, P. Kumar Das, I. Vobornik, X.G. Wan, and D.L. Feng, Signature of Strong Spin-Orbital Coupling in the Large Nonsaturating Magnetoresistance Material WTe$_2$, \href{https://doi.org/10.1103/PhysRevLett.115.166601}{Phys. Rev. Lett. \textbf{115}, 166601 (2015).}

\bibitem{PdTaS2}S. Ni, M. Zhou, Z. Lin, B. Ruan, Z. Li, Z. Zou, Z. Xu, and Z.-an Ren, Crystal growth, superconductivity, and charge density wave of pristine and Pd-intercalated 2\textit{H}-TaS$_2$, \href{https://doi.org/10.1103/PhysRevB.108.075103}{Phys. Rev. B \textbf{108}, 075103 (2023).}

\bibitem{LiTaS2}T. Agarwal, C. Patra, A. Kataria, R. R. Chowdhury, and R. P. Singh, Quasi-two-dimensional anisotropic superconductivity in Li-intercalated 2\textit{H}-TaS$_2$, \href{https://doi.org/10.1103/PhysRevB.107.174509}{Phys. Rev. B \textbf{107}, 174509 (2023).}

\bibitem{CuTaS2}X. D. Zhu, Y. P. Sun, X. B. Zhu, X. Luo, B. S. Wang, G. Li, Z. R. Yang, W. H. Song, and J. M. Dai, Single crystal growth and characterizations of Cu$_{0.03}$TaS$_2$ superconductors, \href{https://doi.org/10.1016/j.jcrysgro.2008.10.023}{Journal of Crystal Growth \textbf{311}, 218 (2008).}

\bibitem{orb_PHE1}H. Wang, Y.-X. Huang, H. Liu. X. Feng, J. Zhu, W. Wu, C. Xiao, and S. A. Yang, Orbital Origin of the Intrinsic Planar Hall Effect, \href{https://doi.org/10.1103/PhysRevLett.132.056301}{Phys. Rev. Lett. \textbf{132}, 056301 (2024).}

\bibitem{orb_PHE2}K. Ghorai, S. Das, H. Varshney, and A. Agarwal, Planar Hall Effect in Quasi-Two-Dimensional Materials, \href{https://doi.org/10.48550/arXiv.2405.00379}{arXiv:2405.00379}

\bibitem{PbS}W. Wan, Y. Yao, L. Sun, C.-C.g Liu, and F. Zhang, Topological, Valleytronic, and Optical Properties of Monolayer PbS, \href{https://doi.org/10.1002/adma.201604788}{Adv. Mater. \textbf{29}, 1604788 (2017).}

\bibitem{TSC1}L. Fu, and C. L. Kane, Superconducting Proximity Effect and Majorana Fermions at the Surface of a Topological Insulator, \href{https://doi.org/10.1103/PhysRevLett.100.096407}{Phys. Rev. Lett. \textbf{100}, 096407 (2008).}

\bibitem{TSC2}H. Yang, Y.Y. Li, T.T. Liu, D.D. Guan, S.Y. Wang, H. Zheng, C. Liu, L. Fu, and J.F. Jia, Multiple In-Gap States Induced by Topological Surface States in the Superconducting Topological Crystalline Insulator Heterostructure Sn$_{1-x}$Pb$_x$Te-Pb, \href{https://doi.org/10.1103/PhysRevLett.125.136802}{Phys. Rev. Lett. \textbf{125}, 136802 (2020).}

\bibitem{TSC3}B. Rachmilowitz, H. Zhao, H. Li, A. LaFleur, J. Schneeloch, R. Zhong, G. Gu, and I. Zeljkovic, Proximity-induced superconductivity in a topological crystalline insulator, \href{https://doi.org/10.1103/PhysRevB.100.241402}{Phys. Rev. B \textbf{100}, 241402(R) (2019).} 

\bibitem{hc1_formula}] J. Bardeen, L. N. Cooper, and J. R. Schrieffer, Theory of superconductivity, \href{https://doi.org/10.1103/PhysRev.108.1175}{Phys. Rev. \textbf{108}, 1175 (1957).}

\bibitem{bg_model}F. J. Blatt, Physics of Electronic Conduction in Solids (McGraw-Hill, New York, 1968).

\bibitem{2gap_model}A. Gurevich, Enhancement of the upper critical field by nonmagnetic impurities in dirty two-gap superconductors, \href{https://doi.org/10.1103/PhysRevB.67.184515}{Phys. Rev. B \textbf{67}, 184515 (2003).}

\bibitem{TaSeS}C. Patra, T. Agarwal, R. R. Chaudhari, and R. P. Singh, Two-dimensional multigap superconductivity in bulk 2\textit{H}-TaSeS, \href{https://doi.org/10.1103/PhysRevB.106.134515}{Phys. Rev. B \textbf{106}, 134515 (2022).}

\bibitem{mcmillan}W. McMillan, Transition temperature of strong-coupled superconductors, \href{https://doi.org/10.1103/PhysRev.167.331}{Phys. Rev. \textbf{167}, 331 (1968).}

\bibitem{SnNb2Se5}H. Bai, L. Qiao, M. Li, J. Ma, X. Yang, Y. Li, Q. Tao, and Z. A. Xu, Multi-band Superconductivity in a misfit layered compound
(SnSe)$_{1.16}$(NbSe$_2$)$_2$, \href{https://doi.org/10.1088/2053-1591/ab60a8}{Mater. Res. Express \textbf{7}, 016002 (2020).}

\bibitem{SnTi2Se5}Y. J. Song, M. J. Kim, W. G. Jung, B.J. Kim, and J.S. Rhyee, Superconducting properties of the misfit-layer compound (SnSe)$_{1.18}$(TiSe$_2$)$_2$, \href{ https://doi.org/10.1002/pssb.201552757}{Phys. Status Solidi B 253, 1517 (2016).}

\bibitem{SnNbSe3}H. Bai, X. Yang, Y. Liu1 M. Zhang, M. Wang, Y. Li, J. Ma, Q.Tao, Y. Xie, G.H. Cao, and Z.A. Xu, Superconductivity in a misfit layered compound (SnSe)$_{1.16}$(NbSe$_2$), \href{https://doi.org/10.1088/1361-648X/aad575}{J. Phys.: Condens. Matter \textbf{30}, 355701 (2018).}

\bibitem{2gap_SH_1}C. L. Huang, J.-Y. Lin, C. P. Sun, T. K. Lee, J. D. Kim, E. M. Choi, S. I. Lee, and H. D. Yang, Comparative analysis of specific heat of YNi$_2$B$_2$C using nodal and two-gap models, \href{https://doi.org/10.1103/PhysRevB.73.012502}{Phys. Rev. B \textbf{73}, 012502 (2006).}

\bibitem{2gap_SH_2}F. Bouquet, Y. Wang, R. A. Fisher, D. G. Hinks, J. D. Jorgensen, A. Junod, and N. E. Phillips, Phenomenological two-gap model for the specific heat of MgB$_2$, \href{https://doi.org/10.1209/epl/i2001-00598-7}{Europhys. Lett., \textbf{56} (6), 856 (2001).} 
\end{thebibliography}
\end{document}